# Towards intense single-digit attosecond pulses with a 100-mJ-class mid-infrared sub-cycle laser

Kaito Nishimiya,[1] Rambabu Rajpoot,[1] Eiji J Takahashi[1]*

[1] Extreme Photonics Research Team, RIKEN Center for Advanced Photonics, RIKEN, 2-1 Hirosawa, Wako, Saitama 351-0198 Japan
* Corresponding author: ejtak@riken.jp

**Abstract**

The duration of isolated attosecond pulses created via high-order harmonic generation is determined by the number of optical cycles in the driving laser. Achieving shorter attosecond soft X-ray pulses requires minimizing the number of cycles while maintaining a high pulse energy. Here, we demonstrate a carrier-envelope-phase-stable, 100-mJ-class sub-cycle mid-infrared laser that produces a supercontinuum coherent soft-X-ray with unprecedented bandwidth. The system delivers 50-mJ, 6.7-fs (0.88-cycle) pulses at a center wavelength of 2.26 μm - over two orders of magnitude more energetic than any previous sub-cycle laser. We applied the system to high-order harmonic generation and compared the results to simulations based on the three-dimensional time-dependent Schrödinger equation to identify unique features of sub-cycle lasers. This work represents a decisive step toward high-energy half-cycle lasers and high-energy single-digit attosecond soft X-ray pulses that can be used to probe matter and light–matter interactions at previously inaccessible temporal resolutions.

**Introduction**

Recent advances in near-infrared (NIR) to mid-infrared (MIR) laser technologies have enabled the development of lasers capable of producing relativistic intensities even in laboratory environments (1) that can reach up to the petawatt level (2). Such lasers have been employed to drive carrier envelope phase (CEP)-dependent strong-field phenomena including isolated attosecond pulses (IAPs) via gas high-order harmonic generation (HHG) (3, 4), efficient soft X-ray emissions via solid-surface HHG (SHHG) (5, 6), and high-energy electrons via laser wakefield acceleration (LWFA) (7). The light or electrons generated by a strong laser field have been utilized in various applications such as the observation of ultrafast electron motion and as a seed for free-electron lasers (8, 9). In these strong-field processes, the wavelength and number of laser cycles critically influence the properties of the emitted radiation. For gas HHG, the maximum photon energy $E_{cutoff}$ follows the cutoff (CO) law $E_{cutoff} \propto I\lambda^2$, where $I$ is the intensity and $\lambda$ is the wavelength. This means that longer wavelengths are more advantageous (10). Meanwhile, reducing the number of cycles for a pulse is beneficial for broadening the continuum region that can be used as an IAP (11). Currently, the shortest duration of an IAP is several tens of attoseconds, and realizing a continuum spanning more than one octave of the soft X-ray spectrum is required to shorten it further (12, 13). For SHHG, the key parameter is the normalized vector potential $a_0$, which scales with $\lambda\sqrt{I}$. Increasing $a_0$ enhances both the conversion efficiency and maximum photon energy, and realizing few-cycle pulses can help facilitate IAPs via SHHG (14, 15). In LWFA, the generation of high-energy electrons requires $a_0 \gg 1$ (1), and few-cycle pulses allow sub-cycle field-driven processes to be separated from cycle-averaged effects (16). These considerations have motivated the development of driving lasers with a stable CEP, high energy, and few-cycle pulses in the MIR region.

Various approaches have been used to develop high-energy fewer-cycle lasers for strong-field applications. Various high-energy lasers requiring two cycles or less have been developed using optical parametric chirped pulse amplification (OPCPA) (17, 18), dual-chirped optical parametric amplification (DC-OPA) (19), and hollow-core fiber (HCF) compression (20) and have been applied to both gas HHG (21) and SHHG (22). Veisz et al. (23) recently developed a 100-TW sub-two-cycle laser using enhanced OPCPA in the visible region. In contrast, high-energy single- or sub-cycle lasers have been difficult to realize because of challenges regarding dispersion compensation and amplification over more than one octave. Tsai et al. (24) developed a single-cycle laser using cascade filamentation (0.98 mJ, 3.1 fs), and Hwang et al. (25) developed a single-cycle laser using a two-stage compressor combining HCF and filamentation (90 µJ, 2.5 fs) (25). However, these lasers have an energy below 1 mJ, which makes energy scaling difficult because of the fundamental limit (26). Chen et al. (27) demonstrated a single-cycle laser based on conventional OPA, but the energy was at the microjoule level. Our group previously developed the advanced DC-OPA method, which employs heterogeneous nonlinear crystals to produce single-cycle pulses at a wavelength of 2.44 µm and energy of over 50 mJ (28). We then used this method to achieve super continuum gas HHG (29).

On the other hand, high-energy sub-cycle or half-cycle lasers exceeding the millijoule level have not yet been realized because of the difficulty of compensation for exceeding one octave of dispersion, seed generation, and amplification. Lin et al. (30) demonstrated 0.7-cycle pulses using two acousto-optic programmable dispersive filters (AOPDFs) for dispersion compensation and OPA in β-barium borate (BBO) pumped by a specially tuned Ti: sapphire laser, but the pulse energy was limited to 30 µJ (7.4 GW). Several groups (31–35) have developed sub-cycle lasers using multicolor waveform synthesis, but the energy levels were all in the microjoule range, which makes their application to high-intensity physics difficult. Only Yang et al. (35) employed a sub-cycle laser for gas HHG, but the continuous region of HHG was narrower than predicted by calculations, and they did not observe the half-cycle cutoff (HCO) (35). Furthermore, optical waveform synthesizer systems are highly complex, and scaling the energy of sub-cycle pulses further to the 100-mJ class is extremely challenging.

In this study, we overcame the energy-scaling limitations of sub-cycle pulses and successfully developed a 100-mJ-class sub-cycle MIR laser system. We applied this system to HHG and observed a supercontinuum high-order harmonics (HH) spectrum, revealing a unique structure of the sub-cycle laser. In HHG, the system has the capability to produce an ultra-broadband continuous spectrum that enables the generation of intense single-digit attosecond pulses. To realize this, we developed a synthesized seed generator (SSG) to produce the 1.4-octave seed pulse required for sub-cycle operation and applied the heterogeneous nonlinear crystals DCOPA (HNC-DCOPA) method to enable amplification across the full spectral bandwidth. As a result, we achieved a 7.6 TW, 2.26 µm, 50 mJ, 0.88-cycle laser - which, to my knowledge, the world's highest-energy sub-cycle laser. We measured the CEP dependence of HHG with Ar using the laser system and confirmed that it matched simulations based on the three-dimensional time-dependent Schrödinger equation (3D-TDSE). We also observed a clear CO energy and a HCO energy, confirming that the continuum region reached 60% as predicted by the calculation, demonstrating that our developed source is indeed a sub-cycle laser. In the discussion part, we focus on three directions: increasing the energy of this sub-cycle lasers, developing half-cycle lasers, and exploring their applications to single-digit attosecond pulse generation.

**Results**

*System design*

Figure 1 shows the layout of the developed 100-mJ-class sub-cycle MIR laser system. The dispersion-tunable CEP-stable 1.4-octave pulse obtained from a SSG is amplified with its full bandwidth preserved by a three-stage HNC-DCOPA. The spectrum of the laser system includes 2.7 μm, where the absorption of water vapor is significant. Thus, the optical path from the SSG to the compressor is inside a sealed box filled with nitrogen. The HNC-DCOPA output pulse is compressed by a sapphire bulk compressor. For experiments, the pulse duration, beam quality, and CEP stability were evaluated by using a third harmonic generation frequency-resolved optical gating (THG-FROG), MIR camera made of HgCdTe (MCT), and $f$–$2f$ interferometer.

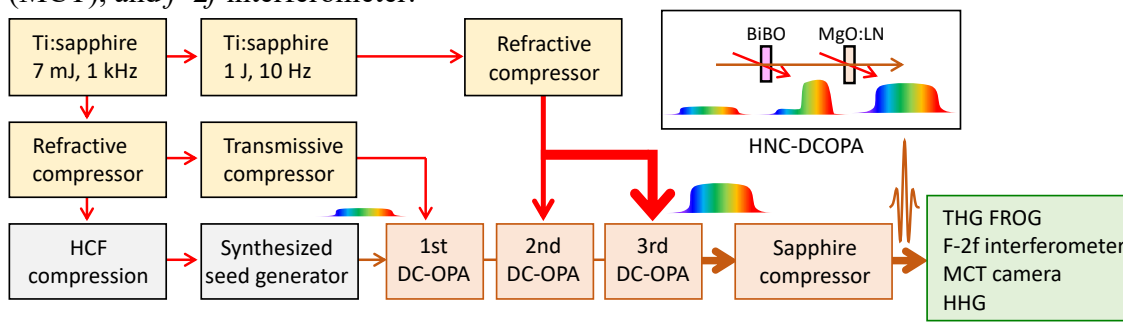

**Fig. 1. Layout of the 100-mJ-class sub-cycle MIR laser system utilizing the HNC-DCOPA method.**

*Synthesized seed generator*

Figure 2(A) shows the configuration of the SSG developed to create a 1.4-octave seed pulse. HCF compression generated a CEP-unstable NIR pulse with a wavelength of 0.5–1.0 μm and energy of 500 μJ, which was focused into the SSG by a concave mirror with a focal length of 3 m. During focusing, the pulse was spatially split by a beam splitter (BS) to generate CEP-stable MIR pulses by intra-pulse difference frequency generation (DFG) using two $BiB_3O_6$ (BiBO) crystals with different nonlinear angles. One optical path generated a seed pulse from the 1.2–2.5 μm region (i.e., shorter-wavelength part), and the other optical path generated a seed pulse from the 2.5–3.2 μm region (i.e., longer-wavelength part). The intra-pulse DFG stabilized the CEPs of both seed pulses even though the CEP of the NIR pulse was not stable (36, 37). Because each seed pulse contained the NIR pulse from HCF compression, the NIR pulse was transmitted by a silicon substrate at Brewster's angle (38). Each BiBO crystal was placed approximately 30 cm upstream of the NIR pulse focus to prevent damage to the crystal and silicon substrate.

In DC-OPA, the seed pulse is compressed after the amplifier, so appropriate dispersion must be applied before the amplifier. Applying as much chirp as possible during compression is advantageous for avoiding crystal damage within the amplifier, reducing the crystal size, and suppressing chirp-matching deviations caused by pulse width changes within the amplifier. This system used a 140-mm-thick sapphire bulk for compression to maximize the pulse width during amplification. Therefore, when pulses were stretched by a specially tuned AOPDF with a low jitter option (DAZZLER/HR45-1100-2200, DAZZLER/HR45-1450-3000), dispersion compensation was required not only for this sapphire bulk but also for the 40-mm-thick CaF$_2$ dichroic mirrors within the amplifier, 15-mm-thick BiBO, 15-mm-thick MgO:LiNbO$_3$ (MgO:LN), and 45-mm-thick TeO$_2$ crystal within the AOPDF. Figure 2(B) shows the group delay required to stretch the pulse width. While the AOPDF offered greater flexibility than other dispersion compensation methods, it could only stretch pulse widths up to approximately 10 ps (shaded region). When the wavelength region was divided into regions of 1.2–2.5 µm and 2.5–3.2 µm, the longer-wavelength part had larger material dispersion than the shorter-wavelength part, which resulted in more than double the pulse width. Therefore, by inserting a 60-mm-thick Si bulk with normal dispersion in the MIR region into the optical path on the longer-wavelength part, and pre-imposing dispersion with the opposite sign, compression was ultimately achieved by using the 140-mm-thick sapphire bulk. When compression was attempted with a 70-mm-thick sapphire bulk and no Si bulk, sufficient compression was not achieved because of the limited tunability of the AOPDF.

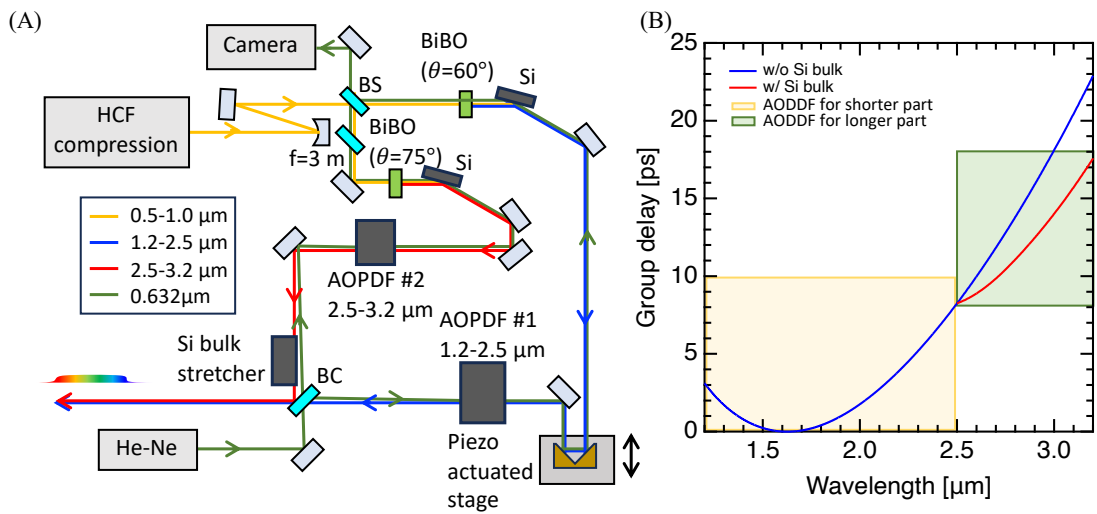

**Fig. 2. SSG for producing a CEP-stable 1.4-octave MIR pulse.** (A) Optical path (BS: beam splitter, BC: beam combiner). (B) Design of the group delay.

The two beams were combined by a custom-made beam combiner (BC). Figure 3(A) shows the reflectivity, reflection group delay dispersion (GDD), and transmission GDD of the BC. A P-polarized short-pass BC was used to minimize the GDD oscillation below 2.5 µm. The delay time between two arms in the SSG was roughly adjusted by moving the manual stage in the shorter-wavelength arm while observing the spectral interference after the SSG. Fine adjustments (<10 fs) were made by measuring the amplified SHG signal and adjusting the delay of one of the AOPDFs in the SSG. Figure 3(B) shows the spectrum of the 1.4-octave seed pulse, which was obtained by adding the spectra from each arm. The stability of the two seed pulses was crucial because each arm had an optical path length of approximately 1 m. We used spatial interference from a He–Ne laser to stabilize the optical path length. The He–Ne laser was incident from the BC side to avoid unnecessary signals from the DCOPA. The zeroth-order diffraction of the AOPDF was used to adjust the incident light because the He–Ne laser cannot pass through the Si bulk. A 1°C change in the temperature of the Si bulk, which is only for the longer-wavelength part, resulted in a relative phase (RP) shift of 12 rad between the two arms. Therefore, the temperature of the Si bulk was controlled to within ±0.004°C. The RP stability of two electric field was evaluated by using the spectral interference of the overlapping part of the MIR seed spectra from each arm. Figure 3 shows the RP stability in terms of the (C) temporal evolution obtained by spectral interferometry with 10-shot averaging, (D) a typical spectrum, (E) the RP stability calculated from (C), and (F) a histogram of (E). The RP demonstrated a stability of 80 mrad (root mean square, rms). Following the work of Cirmi et al. (39), the RP noise from this spectral interference was defined as the quadrature sum of the changes in the physical optical path of the interferometer, the RP changes due to temperature fluctuations, and the CEP noise of each arm. This means that each noise component was 80 mrad, which indicates that the sub-cycle seed pulse had a phase stability of 80 mrad. When the temperature control of the Si bulk was turned off, the spectral interference waveform shifted slowly, which resulted in a slow RP drift (blue line in Fig. 3(E)). On the other hand, when the low jitter option in the AOPDF was turned off and the RP was randomized, the interference spectra canceled each other out, which resulted in a smooth spectrum. These observations support the validity of using the RP to evaluate the phase stability of sub-cycle seed pulses.

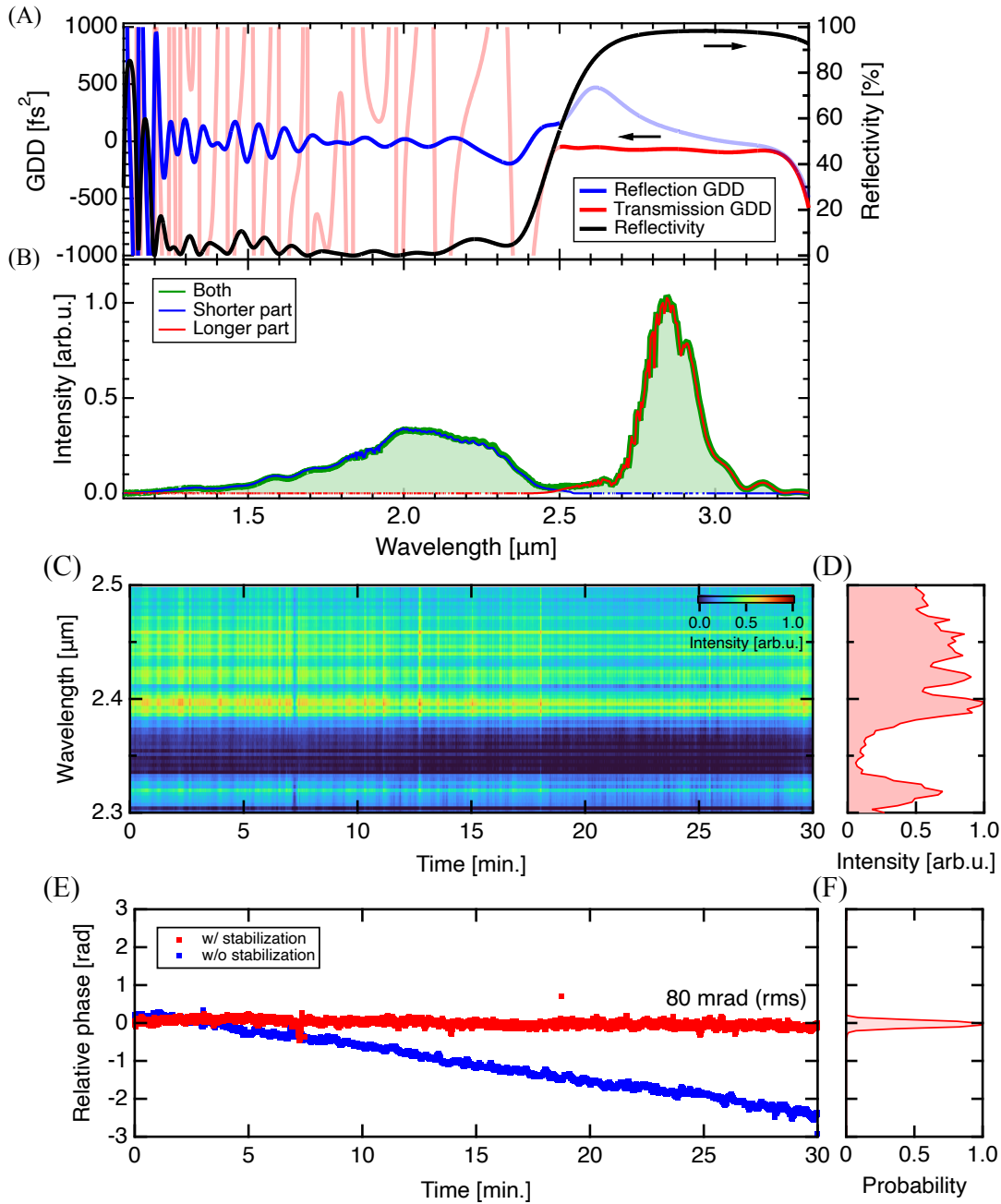

**Fig. 3. Output spectrum and RP stability of the SSG.** (A) Design of the BC. The left axis is the GDD, and the right axis is reflectivity of the BC. (B) Output spectrum of the SSG (C)

Temporal evolution of the interference between the shorter-wavelength (<2.5 μm) and longer-wavelength (>2.5 μm) parts over 0.5 h. (D) Line profile of the interference fringes. (E) RP stability of the interferometer calculated by (C). (F) Histogram of (E).

*Dual-chirped optical parametric amplification with heterogeneous nonlinear crystals*

The 1.4-octave seed pulse was amplified by HNC-DCOPA, which corresponds to our previously developed advanced DC-OPA method (28). DC-OPA amplifies high-energy MIR pulses by applying appropriate chirps to the pump and seed pulses (40), and HNC-DCOPA enables one-octave amplification by cascading two different types of nonlinear crystals (28). We previously discussed that sub-cycle amplification is possible by using three nonlinear crystals with a pump pulse wavelength of 0.76–0.84 μm (28). In this study, we only used two crystals for the amplification of sub-cycle pulses and instead focused on optimizing the nonlinear crystal angle and noncolinear angle of the crystals as well as the wavelength of the pump pulse. Figure 4(A) shows the chirp matching, which is crucial for the development of sub-cycle lasers. The white line indicates the temporal overlap when the seed pulse was chirped by inverse dispersion of a 140-mm-thick sapphire bulk, and the pump pulse had a second-order dispersion of $4.8 \times 10^4$ fs$^2$ with an appropriate delay. Figure 4(B) shows the chirp matching intensity on the white line in Fig. 4(A) and the absorption spectra of 4-mm-thick BiBO and MgO:LN (41–44), which indicates that the optical axis and noncolinear angle of the crystals could be adjusted to amplify wavelengths of 1.2–2.3 μm with BiBO ($\theta = 10.95°$, $\alpha = 0.2°$) and 2.3–3.2 μm with MgO:LN ($\theta = 48.4°$, $\alpha = 1.6°$). In this design, the wavelength of the pump pulse was concentrated in the 0.76–0.82 μm range, which can be amplified by a Ti:sapphire crystal, and efficient amplification of a 1.4-octave bandwidth was achieved by adjusting the wavelength of the pump pulse within this range. MgO:LN has an absorption peak at around 2.7 μm because of incorporated hydrogen ions, but this can be reduced by annealing during the crystal growth process (45). Here, the longer-wavelength part of the 100-mJ-class sub-cycle MIR laser system was limited by the absorption of BiBO. If a laser system can be designed without BiBO, it would be possible to reduce the number of cycles further.

HNC-DCOPA comprised three stages. A 1-kHz pump amplified the first stage while a 10-Hz stage amplified the second and third stages. The chirp of the pump pulse was adjusted to 15 ps, which was confirmed by SHG-FROG based on the changes in the grating distance of the transmitted compressor for the first stage and refractive compressor for the second and third stages. Each stage contained one type-I BiBO crystal and one type-II MgO:LN crystal to facilitate amplification while maintaining a continuum of 1.4 octaves. Figure 4(C)–(E) shows the evolution in the amplification from the first to third stages. Table 1 summarizes the crystal thickness, pump energy, intensity, and output energy for each stage. The output energy was 8.5 μJ after the first stage, 2.8 mJ after the second stage, and 50 mJ after the third stage. The energy after the second stage was reduced by approximately half owing to spatial filtering. In the first stage, because the seed and pump pulses were focused on a nonlinear crystal and amplified, the beam size of the seed pulse from the longer-wavelength part was large. Therefore, we used 2.5 times the energy and 1.5 times the beam size on the longer-wavelength part to amplify the seed pulse while maintaining a constant intensity. To achieve a sub-cycle pulse, the intensity on the longer-wavelength part needed to be increased, but the quantum efficiency on the longer-wavelength part was low. Therefore, 60% of the pump pulse in the second stage and 70% of the pump pulse in the third stage were used for MgO:LN to enhance the longer-wavelength part (>2.5 μm). The first-stage amplifier consumed less than 5% of the pump pulse, which had a small signal gain. The second and third stages were saturated and exceeded 20% of the pump pulse, which resulted in a final conversion efficiency of 6.3%.

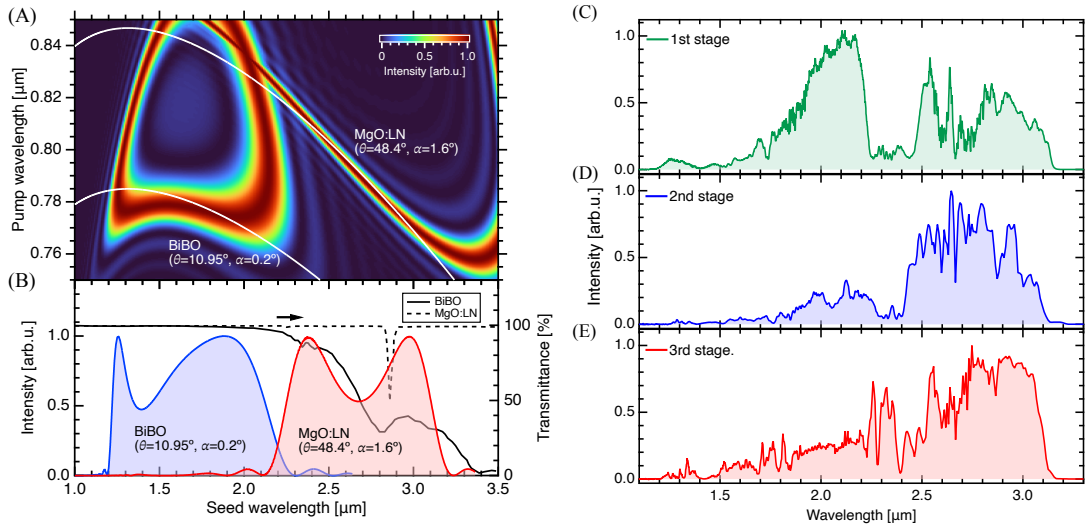

**Fig. 4. Amplification by HNC-DCOPA for the sub-cycle laser.** (A) Chirp matching. (B) Intensity on the white line and absorption spectra of BiBO (41, 42) and MgO:LN (43, 44). Amplified spectra of HNC-DCOPA in the (C) first, (D) second, and (E) third stages.

| Stage \\ Parameter | 1st amp BiBO | 1st amp MgO:LN | 2nd amp BiBO | 2nd amp MgO:LN | 3rd amp BiBO | 3rd amp MgO:LN |
|---|---|---|---|---|---|---|
| Pump energy [mJ] | 0.2 | 0.49 | 24 | 35 | 210 | 490 |
| Pump intensity [GW/cm$^2$] | 17.6 | 17.2 | 9 | 13 | 8.6 | 25 |
| Output energy [mJ] | 0.0046 | 0.0039 | 0.7 | 2.1 | 14 | 36 |
| Crystal Thickness [mm] | 6 | 6 | 5 | 5 | 4 | 4 |

**Table 1. Parameters for each stage of HNC-DCOPA.**

*Characterization of the sub-cycle pulse*

The spectral phase was adjusted by using custom-built software that separated the spectrum within 1.4 octaves into approximately 15 regions. Then the second-, third-, and fourth-order dispersions were determined independently, and the phases were smoothly connected with two types of AOPDFs. We used SHG-FROG for rough adjustment of the spectral phase and THG-FROG for its optimization and measurement. Figure 5(A) shows the measured THG-FROG trace, and Fig. 5(B) is the reconstructed FROG trace, which had an error of 0.86% on a 256 × 256 grid. Figure 5(C) shows the reconstructed spectral intensity, spectral phase, and directly measured spectral intensity obtained by the scanning-type spectrometer. The spectral phase was nearly flat in the range of 1.2–3.2 μm, and the reconstructed and measured spectral intensities showed good agreement. Figure 5(D) shows the time waveform calculated from Fig. 5(C). The full width at half maximum (FWHM) was 6.7 fs, and the number of cycles calculated from the center of gravity of 2.26 μm was 0.88 cycles. This indicates that the 100-mJ-class sub-cycle MIR laser system successfully achieved an output energy of 50 mJ and output power of 7.6 TW.

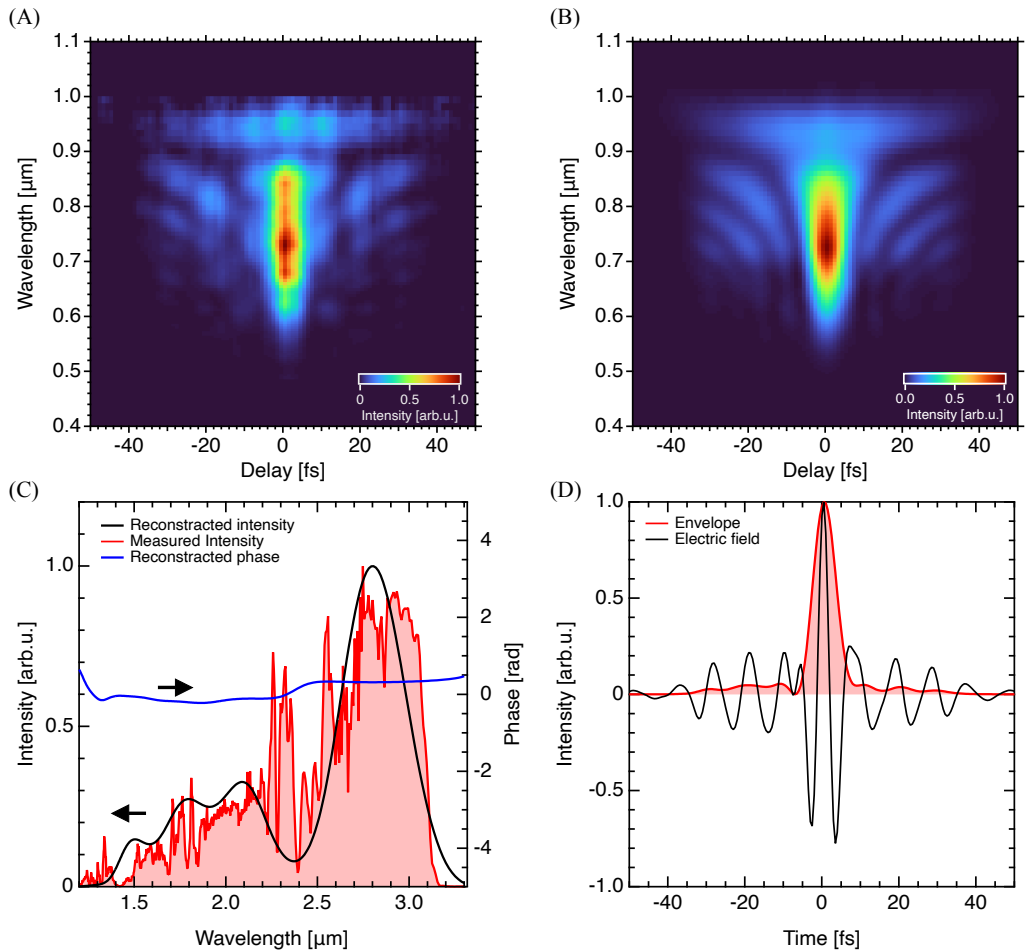

**Fig. 5. Characterization of the amplified pulse with THG FROG.** (A) Measured and (B) reconstructed THG-FROG traces. (C) Reconstructed spectral intensity, spectral phase, and directly measured spectral intensity. (D) Reconstructed electric field and envelope.

We used an f–2f interferometer to evaluate the stability of the CEP after amplification. Part of the compressed pulse was focused into a BBO crystal for SHG, and the intensity ratio between the second harmonic and fundamental components was adjusted by a polarizer. To observe the interference, a delay time between the shorter- and longer-wavelength parts was intentionally generated by the AOPDF. Figure 6 shows the (A) interference spectrum measured over 0.5 h, (B) a typical interference spectrum, (C) the stability calculated from the measured interference spectrum, and (D) a stability histogram. The stability of the CEP was measured as 187 mrad (rms). This measurement method was applied to evaluating both the stability of the CEP and RPs of the two seed pulses in the SSG, and the results indicated that both were sufficiently stable.

Next, we evaluated the beam quality. Because many reflective concave and convex mirrors were used in the propagation path of the 1.4-octave seed pulse to prevent spatial chirp and different crystals were used for seed generation, dispersion compensation, and amplification below and above 2.5 μm, we evaluated the beam quality after amplification for wavelengths below and above 2.5 μm. Figure 6 shows the beam qualities of the (E) shorter-wavelength part and (F) longer-wavelength part after three amplification stages. Even after three stages, the beam quality at the focal point was almost Gaussian with $M^2$ being around 1. The wavelengths used for fitting $M^2$ were 1.8 and 2.7 μm for the shorter- and longer-wavelength parts, respectively. $M^2$ was sometimes less than 1 because the beam before being focused was close to a flat top, but the formula for deriving $M^2$ assumes perfect Gaussian beam propagation.

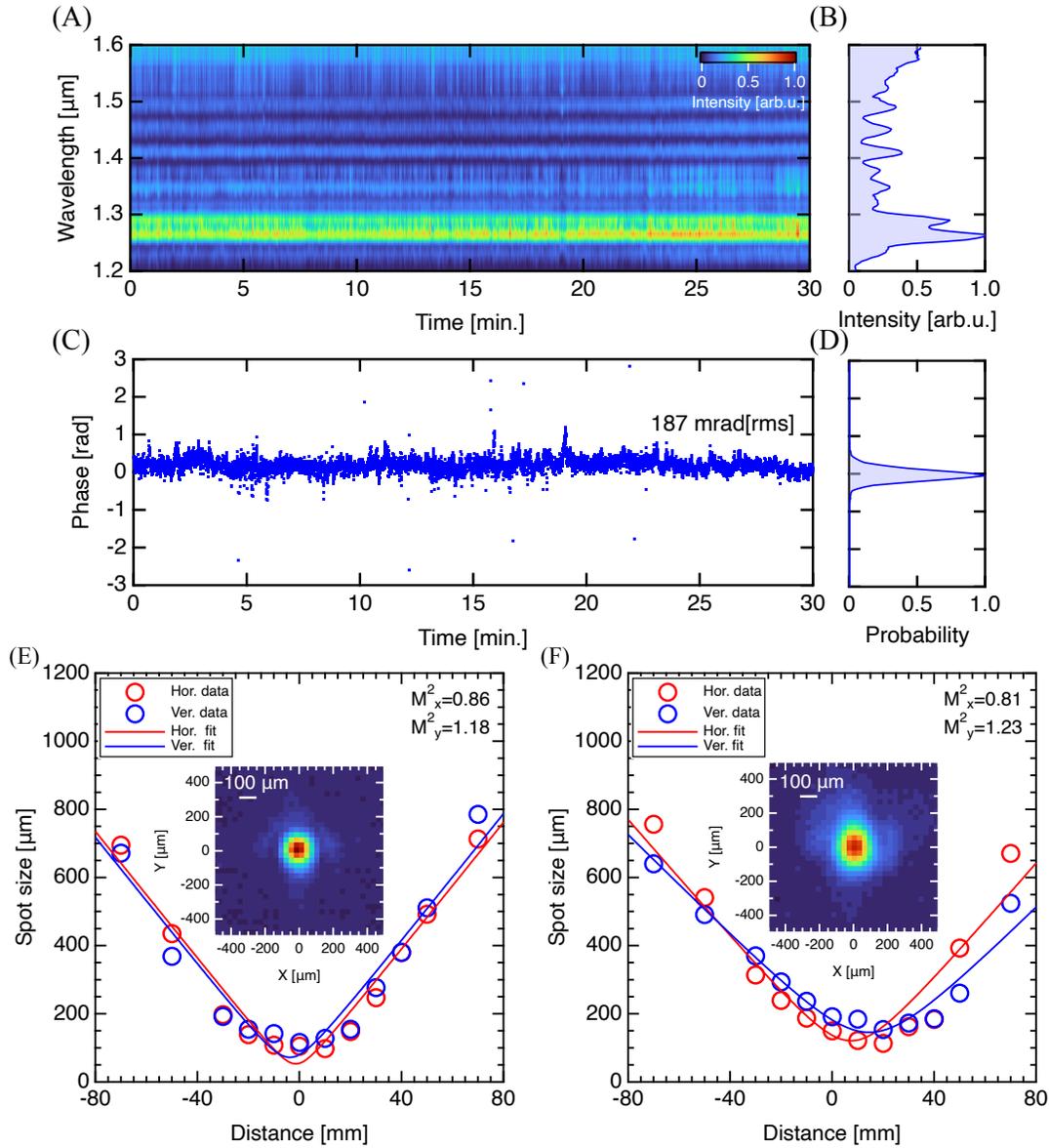

**Fig. 6 CEP stability and beam quality of the sub-cycle laser.** (A) 0.5-h measurement of the interference spectrum with the f–2f interferometer. (B) Typical interference spectrum. (C) CEP stability calculated from (A). (D) CEP stability histogram from (C). (E) beam quality of the shorter-wavelength part (<2.5 μm) (f) Beam quality of the longer-wavelength part (<2.5 μm). The inserts indicate the beam quality at the focal point.

*High-order harmonic generation by a sub-cycle pulse*

There is a correlation between the number of cycles for a laser pulse and the continuous region of HH, and we (29) previously achieved a 40% continuous region with a 1.05-cycle laser (29). Our 100-mJ-class sub-cycle MIR laser system achieved a 0.88-cycle pulse, which according to the CO law means that it should be able to achieve a 60% continuous region for HH. We performed HHG experiments by focusing the developed laser into an Ar gas cell in a vacuum chamber by using an off-axis parabolic mirror with a focal length of 5 m. We used a double-cell configuration with a backing pressure set to 0.5 atm (46). Based on calculations from previous studies, the effective pressure in the interaction region was expected to be less than 3% of the backing pressure at approximately 11 Torr. The focused intensity was adjusted by using an iris before the parabolic mirror to $1.9 \times 10^{14}$ W/cm$^2$, and the focal point was set near the entrance of the gas cell so that phase matching could be satisfied with only a short trajectory (47). The generated HH were transmitted through a 200-nm Zr filter, and the spectra were measured by using a spectrometer with a 1200-groove flat-field grating. The CEP was swept by simultaneously changing the CEPs of the two AOPDFs in the SSG. Figure 7(A) shows the measured CEP dependence of the HH while Fig. 7(B) shows the calculated CEP dependence based on 3D-TDSE using the single-active-electron approximation, which selects only the short trajectory. The CEP dependence of the HH was clearly observed in the experiments, which indicates that a CEP-stable sub-cycle pulse was achieved. The bottom axis of Fig. 7(B) shows the absolute CEP value, and the experimental CEP value was determined by comparing Figs. 7(A) and (B). The incident electric field of the 3D-TDSE was taken from Fig. 5(D) and had an intensity of $1.4 \times 10^{14}$ W/cm$^2$. The dip in Fig. 7(C) at 100 eV is the absorption of Si by the X-ray charge-coupled device (X-CCD) camera used for measurements. The blue lines in Fig. 7(A) and (B) show the changes in CO and HCO. The features of the experimental HH, especially the CO and HCO curves, showed reasonable agreement with the 3D-TDSE results, which confirms that a sub-cycle laser was achieved.

Owing to the sub-cycle nature of the driver pulse, the high-order harmonic signals needed to be evaluated in terms of the electric field intensity rather than the envelope intensity, especially for HH above HCO. For example, the CO energy was approximately 150 eV at a CEP of $1.8\pi$ rad. Decreasing the relative CEP increased the CO energy but decreased the high-order harmonic intensity. The CO energy reached a maximum at a CEP of $\pi$ rad after which the intensity dramatically decreased and returned to 150 eV. The CO energy was determined by the kinetic energy gained from the oscillating electric field by electrons ionized by the electric field at the half-cycle before the moment when the electric field intensity is highest (10). When the CEP was slightly shifted from 0 rad (cosine shape) to either the positive or negative side, there existed CEP values that yielded the same CO energy. However, for sub-cycle laser pulses, the number of tunnel-ionized electrons contributing to the CO energy differed substantially at each CEP, which resulted in a large variation in the HH intensity. For example, changing the CEP by $0.1\pi$ rad led to a ~10% change in the electric field intensity at the ionization time. According to the Ammosov–Delone–Krainov (ADK) theory on tunnel ionization (48), this reduces the number of electrons contributing to the CO energy by more than an order of magnitude. Therefore, both the experiments and theoretical analysis indicated that high-order harmonic signals were no longer observed when the CEP was changed from 0 rad (cos shape) to the negative side.

The vertical axis in Fig. 7(C) shows the photon flux calibrated with the Zr filter transmittance, the quantum efficiency of the X-ray CCD camera, and diffraction efficiency of the grating (49). The CO and HCO energies changed significantly when the CEP is altered. The left insert shows an enlarged view of the low-photon-energy side when CEP = $\pi/4$. A comb structure was observed at the HCO frequency, and the fundamental wavelength calculated from this comb was approximately 2.25 µm, which is consistent with the center of gravity wavelength of the laser. The peak at around 83 eV was attributed to second-order diffraction from the grating, and its location on the higher-energy side of the comb structure confirmed that the continuous region spanned more than one octave. A continuous region was observed from 75 eV to 180 eV, which is higher than the HCO photon energy, and the continuous region accounted for approximately 58% of the total photon energy range, which is consistent with the calculations. The spectral shape in the photon energy region above

HCO comprised a relatively flat plateau region and a peaked cutoff region, which clearly reflects a single-atom response and can be explained by the three-step model (10, 50) and the use of a half-cycle driving laser. Because both the short and long trajectories contributed to the recombination process in the CO region, electrons ionized within the longer time window contributed to the recombination process occurring over a narrower photon energy range in the CO region than in the plateau region. For the short trajectories, the ionization timing for higher photon energies resulted in a higher instantaneous electric field intensity than for lower photon energies, which allowed more electrons to participate in the recombination process. On the other hand, when the free electrons were accelerated by the oscillating electric field over a long time, the number of electrons available for recombination decreased because quantum diffusion. These effects counteracted each other, which explains the relatively flat plateau region and peaked cutoff region (51).

The energy in the continuum region calculated from the photon number observed by the X-CCD camera was approximately 0.73 nJ if the energy loss in the slit (200 µm) is considered. Because the absorption length under these experimental conditions was approximately 30 mm, the absorption-limited condition on the phase matching (52) was not satisfied with the current gas cell length of 50 mm. The beam radius at the focal point was 500 µm, and the Rayleigh length is 35 cm. Thus, extending the gas cell length to 150 mm would enable a threefold increase in the output energy. The Fourier transform-limited (FTL) pulse width in the obvious continuum region of this spectrum (i.e., 75–180 eV) was 34.7 as. Li et al. (13) conducted pulse width measurements using a photon energy of 100–300 eV but achieved a pulse width of only 53 as, and the pulse energy was approximately 10 pJ because of insufficient dispersion compensation for IAPs. This is because dispersion compensation for HH is difficult above 250 eV when a single filter is used. In this study, we demonstrated supercontinuum HHG below 200 eV where dispersion compensation can be achieved with a proper metal filter. Thanks to this feature, we were able to generate shorter IAPs with an energy that is more than two orders of magnitude higher, which is suitable for single-cycle soft X-rays.

The bandwidth of the generated attosecond pulses exceeded one octave, which is significant. For example, attosecond pulses that span more than one octave can be used to observe the interference between photoelectrons due to two-photon absorption on the low-energy side of the spectrum and photoelectrons due to one-photon absorption on the high-energy side. This would make it possible to observe the relative CEP of the high-order harmonic in the time domain and carrier envelop offset (CEO) in the frequency domain. Although several optical frequency combs have been demonstrated, particularly in the extreme-ultraviolet region (53, 54), direct measurement of the CEO has not yet been demonstrated. Only light in the visible region has been transcribed. This is because the peak intensity cannot be increased as in the visible region, which makes it impossible to generate more than one octave as required for CEO measurements. In contrast, our 100-mJ-class sub-cycle MIR laser system can achieve HHG over a bandwidth of more than one octave, which opens the possibility of direct CEO measurement in the soft X-ray region.

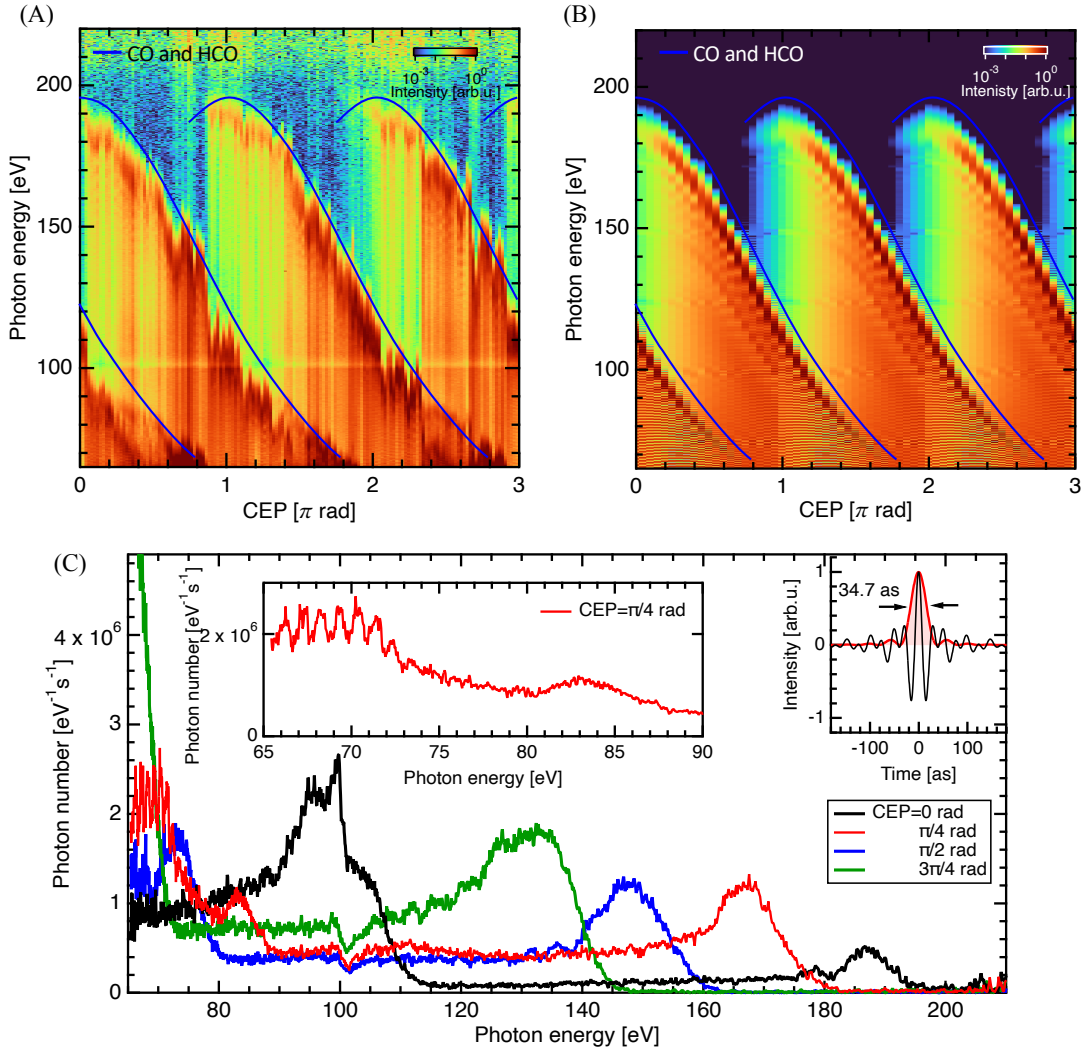

**Fig. 7. High-order harmonic generation by the sub-cycle laser.** (A) Experimental CEP dependence. (B) Calculated CEP dependence. The blue lines correspond to the CO and HCO curves. (C) Line profile of the high-order harmonic spectrum. The left inset shows an enlarged view from 65 eV to 90 eV when CEP is $\pi/4$, and the right inset shows the FTL pulse when the continuum region is extracted.

**Discussion**

Here, we discuss the future prospects of our 100-mJ-class sub-cycle MIR laser system as well as applications to high-intensity physics, and we present a roadmap toward single-digit attosecond pulse generation.

First, the energy of the sub-cycle laser can be expanded from the 100-mJ class to the 1-J class. To increase the output energy, it will be necessary to increase the pulse width and beam size to avoid damage to the nonlinear crystals, as is the case with chirped pulse amplification. Currently, the third-stage amplifier utilizes a crystal size of 20 mm × 20 mm. However, the beam diameter has been adjusted to 15 mm. The current size limit for BiBO and MgO:LN crystals is 20 mm × 20 mm, but 30 mm × 30 mm crystals can also be manufactured. For example, suppose that the thickness of the sapphire bulk for the compressor was extended to 280 mm, which would double the pulse width, and the beam diameter was set to 20 mm for the BiBO crystal and 30 mm for the MgO:LN crystal. In that case, the intensity would decrease by 0.18 times and 0.08 times, which would allow pumping at higher energies. With this setup, using a 10-J-class Ti:sapphire pump laser would allow the output energy to exceed 500 mJ.

Next, the sub-cycle laser may be improved to achieve a half-cycle laser. Half-cycle lasers have previously been achieved by using a synthesizer method to produce an output energy of 320 µJ (32) or two-color filamentation to produce an output energy 420 nJ (55). We suggest that a half-cycle laser with an output energy of up to 100 mJ can be achieved by combining the SSG with HNC-DCOPA in the longer-wavelength part. The wavelength range of the current laser system has a wavelength range of 1.2–3.2 µm, whose upper bound is due to the strong absorption above 3.2 µm of the BiBO crystal used for amplification of the shorter-wavelength part (41, 42). Therefore, HNC-DCOPA can be upgraded with two MgO:LN and KTiOAsO$_4$ (KTA) crystals to develop a half-cycle laser in the 1.2–4.0 µm region. Figure 8(A) shows the chirp matching with two type-I MgO:LN crystals ($\theta = 47.9°$, $\alpha = 0.2°$; $\theta = 48.8°$, $\alpha = 1.5°$), and one type-I KTA crystal ($\theta = 34.96°$, $\alpha = 0.2°$). The white line indicates the temporal overlap when the seed pulse is chirped by inverse dispersion of a 140-mm sapphire bulk and the pump pulse has a second-order dispersion of $7.0 \times 10^4$ fs$^2$. Figure 8(B) shows the chirp matching intensity for the white line in Fig. 8(A) and the absorption spectra of the MgO:LN and KTA crystals with thicknesses of 4 mm (43, 44, 56). By adjusting the nonlinear angle and noncollinear angle, we found that we could amplify the 1.2–2.8 µm region with the MgO:LN crystals and 2.5–4.0 µm region with the KTA crystal. The seed pulse for these lasers can be generated by using the SSG developed in this study. The output of the Ti:sapphire laser system is also compressed by an HCF and then input to the SSG. For seed generation, an intra-pulse DFG with a BiBO crystal is used for the shorter-wavelength part covering 1.2–2.5 µm, and the cascaded nonlinear process from Yin et al. (57) can be used for the longer-wavelength part covering 2.5–4 µm. For dispersion compensation, a Si bulk and AOPDF are used together to account for the sapphire bulk used for compression. Specifically, the AOPDF for the shorter-wavelength part (DAZZLER/HR45-1100-2200) and AOPDF for longer-wavelength part (DAZZLER/WB45-2000-3700) are tuned to cover 1.2–2.5 µm and 2.5–4.0 µm, respectively. This enables seed generation, dispersion compensation, and amplification across the entire 1.2–4.0 µm range, which would enable the development of a 100-mJ-class half-cycle laser with a center of gravity wavelength of 2.59 µm as shown in Fig. 8(C) along with a duration of 5.13 fs and pulse of 0.59 cycles as shown in Fig. 8(D).

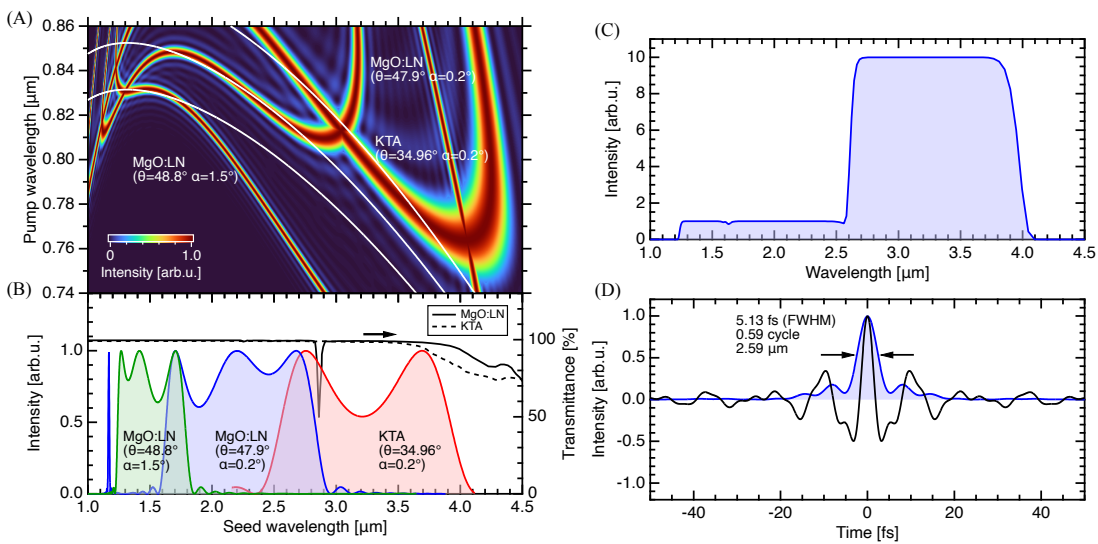

**Fig. 8. Proposed design of a half-cycle laser.** (A) Chirp matching. (B) Intensity on the white line in (A) and absorption spectra of MgO:LN (43, 44) and KTA (56) crystals. (C) Spectral intensity and (D) time waveform of the half-cycle laser.

One of the potential applications of a sub-cycle laser is to generate single-digit attosecond pulses of soft X-rays. The shortest width for attosecond pulses is currently 43 as (12), and a broader bandwidth is necessary to further reduce the pulse width. In this study, we generated broadband attosecond pulses with a 58% bandwidth of 75–180 eV using Ar gas. If we apply the developed laser system to HHG using He gas as a nonlinear medium, the cutoff energy would reach 700 eV (58), which should extend the continuum region to 300–700 eV and make up 60% of the wavelength range. With a FTL pulse width of 9.5 as, this sub-cycle laser would provide sufficient bandwidth for single-digit attosecond pulse generation. Furthermore, applying the half-cycle laser proposed in Fig. 8 to HHG with He gas would achieve a cutoff energy of 1 keV and extend the continuum region to 200–1000 eV range and 80% of the wavelength range, which would generate single-digit attosecond pulses with a duration of 4.9 as.

Compensating for atto-chirp remains an important challenge. Dispersion compensation using a metal filter has been demonstrated to be effective below 250 eV(12, 57). Dispersion compensation has been proposed using fully ionized hydrogen plasma for the water window region (284–530 eV) (59) and unionized molecular hydrogen gas for the region between 530 eV and 1 keV (60), but neither approach has been experimentally demonstrated. In addition, third-order dispersion remains a problem. HHG by the developed sub-cycle laser system and proposed half-cycle laser would facilitate the verification of dispersion compensation in these bands.

Another application of shorter attosecond pulses is SHHG. One advantage of SHHG is that atto-chirp is negligible, which is a large problem for gas HHG (61). Single attosecond pulses can be generated by reducing the number of cycles of the driving laser (21, 62). The developed 100-mJ-class sub-cycle MIR laser system is the only light source capable of generating SHHG using a sub-cycle laser. If the beam is focused to a diameter of 5.5 µm, it can achieve an intensity of $6.3 \times 10^{19}$ W/cm$^2$ and $a_0 \sim 15$ with a cutoff photon energy of 520 eV. Because dispersion compensation would not be required, it would be possible to generate a single-digit attosecond pulse of 9.1 as by using the bandwidth from 100 eV to 500 eV.

**Methods**

*Pump for HNC-DCOPA*

A custom-made Ti:sapphire chirped pulse amplification system was used as the frontend laser, and the chirped pulse was amplified to 7 mJ at a repetition rate of 1 kHz. Of this energy, 4 mJ was used to synthesize seed and pump pulses for the first stage of HNC-DCOPA, and the remaining 3 mJ was amplified to 1 J at 10 Hz for the pump pulses of the second and third stages. To generate the 1.4-octave seed pulse, the 800-nm pulse from the Ti:sapphire chirp pulse amplification system was first broadened by self-phase modulation. The light from the Ti:sapphire laser (1 mJ, 25 fs) was focused into a Ar-filled HCF by using a 4-m focal length lens. A stretched fiber was used for the HCF, and the pressure-gradient HCF method was used with gas flowing from the output end. The incident energy was adjusted to 800 µJ to achieve a throughput of approximately 80%. This light was collimated with a concave mirror, and dispersion was compensated by using chirped mirror pairs and a wedge pair, which resulted in compression to 5.5 fs. The central 500-µJ part of the beam was used for the SSG. The light (1 kHz, 2 mJ) was used as the pump pulse for the first stage of amplification, and negative dispersion was imparted by using a pulse compressor with a transmission grating pair to stretch the pulse width to 15 ps (FWHM). The remaining chirped light (1 kHz, 3 mJ) was amplified to 1 J by an additional 10-Hz two-stage multi-pass amplifier, and negative dispersion was imparted by a vacuum pulse compressor that stretched the pulse width to 15 ps.

The optical configuration of the pump pulse for HNC-DCOPA was designed to achieve the cleanest possible spatial profile. Because the light in the first stage had low energy, the pump pulse was focused and amplified within the crystal. Therefore, image transfer was performed after the 1-kHz transmission compressor, and the beam quality at the focal point was designed to match a Gaussian beam profile. The pulse energy was changed by adjusting the aperture of an iris placed before the focusing lens. In the second and third stages, the light was down-collimated rather than focused because of its high energy. The output light from the 10-Hz compressor was a flat-top beam with a full width at half maximum of approximately 70 mm, and the beam diameter was adjusted by using a two-stage Kepler-type down-collimator. Because the beam quality deteriorated with propagation after Kepler-type down-collimation, a second lens was placed approximately 10 cm from the crystal.

*Measurement of the pulse duration*

A beam-splitting symmetric Michelson interferometer was used with one of the roof mirrors controlled by a piezo-driven stage. The light from each arm was focused onto a nonlinear crystal by using a concave mirror with a focal length of 20 cm. A 10-µm-thick BBO crystal was used for SHG-FROG while a 200-nm-thick $Si_3N_4$ crystal was used for THG-FROG. In the SHG-FROG measurements, the spectral bandwidth of the resulting SHG could not be fully measured by a spectrometer equipped with a Si-based detector alone. Therefore, the second-harmonic light was split by using a beam splitter, and one part was simultaneously measured by using a NIR spectrometer equipped with an InGaAs-based detector to evaluate the pulse width. THG-FROG could be measured simply by removing the beam splitter and changing the nonlinear crystal from BBO to $Si_3N_4$.

**Acknowledgments**

We thank Dr. Y. Nabekawa for the helpful comments on drafting this manuscript. K.N. acknowledges the Special Postdoctoral Researcher's Program of RIKEN.

**Funding:** We acknowledge financial support from the Ministry of Education, Culture, Sports, Science and Technology of Japan (MEXT) Quantum Leap Flagship Program (Q-LEAP) (grant no. JP-MXS0118068681).

**Author contributions:** K.N. designed the experimental setup, performed the experiments, and analyzed the experimental data. R.R. performed the numerical calculations of HHG. K.N. and E.J.T. discussed the experimental results and wrote the manuscript. E.J.T. initiated and supervised this project as a whole.

**Competing interests:** The authors declare that they have no competing interests.

**Data and materials availability:** All data needed to evaluate the conclusions in the paper are present in the paper and/or the Supplementary Materials. Additional data related to this paper may be requested from the authors.